\documentclass[11pt,a4paper,english]{article}
\usepackage[right=0.75in, left=0.75in, top=1.20in, bottom=1.20in]{geometry}
\usepackage[utf8]{inputenc}
\usepackage[T1]{fontenc}
\usepackage[english]{babel}
\usepackage{enumitem}
\usepackage{xcolor}   
\usepackage{amsmath,amsfonts,amssymb,amsthm}
\usepackage{graphicx}
\usepackage{mathrsfs}
\usepackage[affil-it,auth-sc]{authblk}
\usepackage{pgf,tikz}
\usetikzlibrary{decorations.pathmorphing,patterns}
\usetikzlibrary{shapes.geometric}
\usetikzlibrary{arrows}
\usetikzlibrary{snakes}
\usepackage{siunitx}
\usepackage{gensymb}
\usepackage{sectsty}
\graphicspath{{Images/}}

\newcommand{\defeq}{\mathrel{\mathop:}=}

\newcommand*{\diff}{\mathop{}\!\mathrm{d}}
\newcommand{\dd}{\diff}

\newcommand{\bege}{\begin{equation}}
\newcommand{\eend}{\end{equation}}
\newcommand {\Ec} {\mathscr{E}}
\newcommand{\eps}{\varepsilon}
\newcommand{\epso}{\eps_{\rm o}}
\newcommand{\epsodot}{\dot\eps_{\rm o}}
\newcommand{\epsomax}{\epso^{\rm max}}
\newcommand{\keps}{\alpha}
\newcommand{\td}{t_d}
\newcommand{\tc}{t_c}
\newcommand{\tcc}{t_c^*}

\newcommand{\Tmin}{T_\mathrm{min}}
\newcommand{\Tmax}{T_\mathrm{max}}
\newcommand{\up} {u^\prime}
\newcommand{\ud} {\dot u}
\newcommand{\Tp} {T^\prime}

\newcommand {\ub} {\mathbf{u}}
\newcommand {\vb} {\mathbf{v}}
\newcommand{\A}{\mathcal{A}}
\newcommand{\Rc}{\mathcal{R}}

\DeclareMathOperator{\Diss}{Diss}
\DeclareMathOperator{\diss}{diss}
\DeclareMathOperator{\argmin}{argmin}
\providecommand{\abs}[1]{\left\lvert#1\right\rvert} 
\numberwithin{equation}{section}
\sectionfont{\normalsize}
\subsectionfont{\normalsize}

\begin{document}
\author[1,2,\footnote{Corresponding author -- e-mail address:\,\texttt{desimone@sissa.it} -- phone:\,+39\,040\,3787\,455}]{Antonio DeSimone}
\author[1]{Paolo Gidoni}
\author[1]{Giovanni Noselli}
\affil[1]{\small{SISSA - International School for Advanced Studies, via Bonomea 265, 34136 Trieste, Italy.} \smallskip}
\affil[2]{\small{GSSI - Gran Sasso Science Institute, viale Francesco Crispi 7, 67100 L'Aquila, Italy.} \smallskip}
\title{Liquid Crystal Elastomer Strips as Soft Crawlers}
\date{}
\maketitle
\begin{abstract}
\noindent In this paper, we speculate on a possible application of Liquid Crystal Elastomers to the field of soft robotics. In particular, we study a concept for limbless locomotion that is amenable to miniaturisation. For this purpose, we formulate and solve the evolution equations for a strip of nematic elastomer, subject to directional frictional interactions with a flat solid substrate, 
and cyclically actuated by a spatially uniform, time-periodic stimulus (e.g., temperature change). The presence of frictional forces that are sensitive to the direction of sliding  transforms reciprocal, \lq breathing-like' deformations into directed forward motion. We derive formulas quantifying this motion in the case of distributed friction,  by solving a differential inclusion for the displacement field. The simpler case of concentrated frictional interactions at the two ends of the strip is also solved, in order to provide a benchmark to compare the continuously distributed case with a finite-dimensional benchmark. We also provide explicit formulas for the axial force along the crawler body.
\end{abstract}

\section{Introduction}\label{sec:Intro}

Liquid Crystal Elastomers (LCEs) are polymeric materials that can exhibit spontaneous deformations when activated by light, heat or electric fields, thanks to the coupling between rubber elasticity and nematic order (Warner and Terentjev, 2003). Among the possible applications that are being envisaged, some of the most spectacular ones involve their use for locomotion purposes and as motors or soft manipulators (Camacho-Lopez {\it et al.}, 2004; Ikeda {\it et al.}, 2007; Fukunaga {\it et al.}, 2008; van Oosten {\it et al.}, 2008; Sawa {\it et al.}, 2010; Wei and Yu, 2012; Knezevic and Warner, 2013).

In view of their striking properties, LCEs seem to be particularly appealing in the realm of \emph{soft robotics}.  According to this new paradigm in robotic science, inspiration is sought from nature to endow robots with new capabilities in terms of dexterity or adaptability, by exploiting large deformations typical of soft materials. As far as dexterity is concerned, one may think, for example, of the manipulation abilities of an elephant trunk or of an octopus arm. As for adaptability, one may think of the ability of snakes to handle unexpected interactions with unstructured environments and move successfully on uneven terrains by adapting their gait to ground properties that change from place to place in an unpredictable way. Indeed, manipulation and locomotion are among the most intensively investigated applications of current soft robotics research (Hirose, 1993; Trivedi {\it et al.}, 2008;  Kim {\it et al.}, 2013). Recent studies of bio-inspired mechanisms for shape control can be particularly interesting for this kind of applications (Armon {\it et al.}, 2011;  Arroyo {\it et al.}, 2012; Arroyo and DeSimone, 2014). 

Furthermore, LCEs  are mechanically compliant, with elastic moduli comparable to those of skeletal muscles and other biological tissues, unlike the case of hard, metal based active materials. For this reason, they may prove particularly useful in the future development of medical micro-sized machines such as endoscopic robotic capsules. In fact, Stefanini {\it et al.} (2006) argue that active and controllable locomotion is one of the key features which can transform a high-tech miniaturized vision system into a really useful device for endoscopic diagnosis and therapy.

While the compliance and large spontaneous deformations of LCEs make these materials appealing, they raise considerable modelling challenges. Indeed, the strong material and geometric nonlinearities associated with their interesting behaviour  make the problem of predicting and controlling their response to prescribed stimuli quite difficult.

Restricting attention to bio-inspired self-propulsion mechanisms, one can find several studies in the recent literature that have drawn inspiration from earthworm locomotion (Gray and Lissmann, 1938; Quillin, 1999; McNeil, 2003). A common feature in these studies is that traveling contraction waves span the slender, straight body of a model crawler which uses nonlinear, frictional interactions with a flat surface in order to move  (Mahadevan {\it et al.}, 2004; Menciassi {\it et al.}, 2006; Tanaka {\it et al.}, 2012; DeSimone and Tatone, 2012; DeSimone {\it et al.}, 2013; Noselli {\it et al.}, 2013; Gidoni {\it et al.}, 2014; Noselli and DeSimone, 2014).
The nonlinearity of the interactions is exploited to produce spatially modulated stick-slip patterns similar to those observed in earthworms, that stick to the surface in the contracted portions of the body (where larger friction forces are available thanks to the contraction-induced protrusion of \emph{setae}, which enhance surface roughness and traction) and slide in the smoother, non-contracted portion.

Prototypes of soft robotic crawlers have been realised, made of elastomeric bodies with embedded shape memory alloy (SMA) wires. These were selectively activated, via the Joule heating of the SMA actuators, to induce waves of localized contractions (Menciassi {\it et al.}, 2006). LCEs  provide us with the possibility of dispensing with the SMA wires and with the need of driving them with carefully controlled spatio-temporal actuation patterns.
In fact,  a crawler  with a body made of an LCE would undergo large extensions and contractions when subjected to various spatially uniform, time-periodic stimuli (such as light, heat, electric field) directly applied to the LCE. And while producing travelling contraction waves requires complex actuation strategies and stimuli modulated in space and time, we argue that a strip of LCE, simply actuated by spatially uniform stimuli inducing breathing-like deformations, would exhibit net displacements when placed on a \emph{directional surface}, see the review by Hancock {\it et al.} (2012). The key properties of such surfaces, a simple example of which is a patch of hairy skin, is that they offer different resistance to motion along and against the grain. In other words, the force-velocity law characterizing frictional interactions on directional surfaces is not odd. More generally, we use the term \emph{directional frictional interactions} to refer to force-velocity laws of this type, allowing for the possibility that the inclined flexible elements (e.g., hair or bristles) mediating the contact interaction could be attached to the crawling strip, rather than to the substrate. Evidence that  directional frictional interactions can be actually implemented trough the use of `asymmetric feet' can be found in Menciassi {\it et al.} (2006) and Noselli and DeSimone (2014), for example.
It has been already shown in a number of theoretical and experimental studies that, in these circumstances, it is feasible to extract net positional changes out of breathing-like deformation modes (Mahadevan {\it et al.}, 2004; Gidoni {\it et al.}, 2014; Noselli and DeSimone, 2014). Interestingly, this is at odds with what happens in low Reynolds number swimmers where, according to the so-called Scallop Theorem (Purcell, 1977), the net displacement associated with a time-reversible history of shape changes is always zero.

Having argued in favour of the idea of using oscillations in gels for locomotion purposes (an idea which is not new, see, e.g., Yoshida {\it et al.} (1996), Mahadevan {\it et al.} (2004) and Maeda {\it et al.} (2007)), we are now ready to discuss our results. In this paper, we formulate and solve a one-dimensional model for the motion of a strip of nematic LCE subject to  directional frictional interactions and to a prescribed, spatially uniform time-history of spontaneous distortions.  We emphasise that, in this study, we do not assume the shape of the crawler to be known a-priori, and freely prescribed through a given time history (this is the view-point taken in Gidoni {\it et al.} (2014)). Instead, similarly to what is done in Noselli and DeSimone (2014) for a crawler resting on two elastic bristles, the time-evolving configuration of the crawler is an \emph{emergent} property which arises from the coupled nonlinear system consisting of the crawler force-generating mechanism (the time-varying active distortions), its passive elasticity (associated with its extensional elastic stiffness, quantified by the area of its cross section and its Young's modulus), and the external frictional forces.

Our focus in this paper is on the conceptual challenges that our model raises  (in particular: what is the axial force along the crawler body accompanying locomotion? What are the net displacements that can be extracted from a reciprocal actuation strategy producing breathing-like deformations?), rather than on the challenging question of how to actually build a device behaving in accordance with our model (see however the Discussion  section for some further remarks in this direction.)
Our main results are (i)\,the formulation and solution of the equations of motion for a strip of nematic elastomer subject to directional frictional interactions and uniform distortions, which is, to the best of our knowledge, obtained here for the first time, and (ii) explicit formulas for the achievable displacements and for the axial force along the strip.

The rest of the paper is organised as follows. The equations governing the evolution of the LCE crawler are derived in Section 2, and the motility problem is formulated in Section 3. The case of frictional interactions acting only at the crawler extremities is explored in Section 4 as a warm up to the more difficult case of distributed interactions, and to provide a benchmark to contrast the more complex case against a simpler one. The case of distributed friction is considered in Section 5. The Appendix shows how the evolution equations governing the motion of the crawler can be obtained from an incremental minimisation problem.

\section{Governing equations}\label{sec:GovEq}

\begin{figure}[h]
\renewcommand{\figurename}{\small {\bf Fig.}}
\centering
\includegraphics{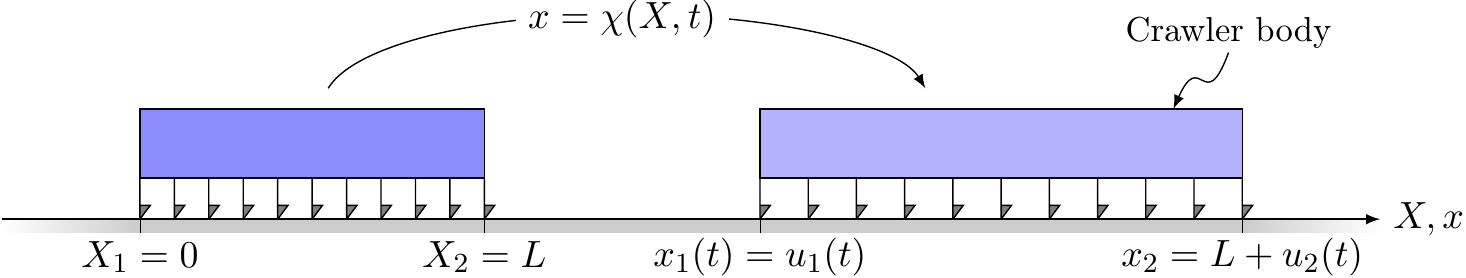}
\caption{\small A sketch of the one-dimensional crawler analysed in this study. The model accounts only for horizontal displacements along the $X$ coordinate, whereas the system exploits directional frictional interactions with a solid substrate either at its extremities (case of interactions only at the extremities) or along its body length (case of distributed interactions, shown in the figure).}
\label{fig:crawlers}
\end{figure}

We consider the model crawler shown in Fig.~\ref{fig:crawlers} and denote the position of its points through a one-to-one function $\chi(X,t)$ mapping the reference configuration $[X_1,X_2]$
(more concretely, $X_1=0$ and $X_2=L$, $L$ being the reference length of the crawler) onto the deformed configuration $[x_1(t),x_2(t)]$ where
\bege
\left\{
\begin{array}{l}
\!\! x_1(t)=\chi(X_1,t)=X_1+u(X_1,t)= u_1(t) , \\[2mm]
\!\! x_2(t)=\chi(X_2,t)=X_2+u(X_2,t)= L+ u_2(t) .
\end{array}
\right .
\eend

Here $u(X,t)$ is the displacement at point $X$ and time $t$ defined by
\bege
u(X,t) = x - X = \chi(X,t) - X ,
\eend
whereas $u_1(t)$ and $u_2(t)$ are the displacements at time $t$ of the two end points. We will denote with primes and dots the partial derivatives with respect to space and time, respectively, according to
\bege
u^\prime(X,t) \defeq \frac{\partial}{\partial X} u(X,t) , \quad \dot{u}(X,t) \defeq \frac{\partial}{\partial t} u(X,t) .
\eend

The body of the crawler is made of a nematic LCE. We assume that its elastic energy is given by
\bege\label{energy}
\Ec(u,t)=\int_0^L \frac K2\left( \eps_u(X,t)-\epso(X,t) \right)^2 \dd X ,
\eend
where
\bege
\eps_u(X,t)=u^\prime(X,t)
\eend
is the strain, $K>0$ is the 1D elastic modulus (with dimension of force since $K=EA$, where $E$ is Young's modulus and A the cross-sectional area), and $\epso(X,t)$ is the spontaneous, or stress-free strain at $X$ and $t$.
We assume that  $-1<\epso<+\infty$ and refer to $\epso$ as the \emph{active distortion}: in analogy with thermal dilatation, it is the spontaneous strain (i.e., the one in the absence of stress) associated with a phase transition. It can model the spontaneous deformation accompanying either the nematic-to-isotropic transition (which can be induced by increasing the temperature past the phase transition temperature, or by irradiation with UV light in the case of photosensitive elastomers), or the isotropic-to-nematic transition induced by cooling a specimen initially in the isotropic state. Alternatively, it can be the spontaneous deformation accompanying a director reorientation in a nematic specimen (say, from perpendicular to parallel to the crawler axis, that can be induced by the application of a suitably oriented electric field).
In LCEs, spontaneous strains can be exceptionally large: the spontaneous extension accompanying the isotropic-to-nematic transition can be as large as $300\%$ (Warner and Terentjev, 2003). For this reason, we put no restrictions on the magnitude of the spontaneous strain, which can be arbitrarily large.

In expression \eqref{energy} we have used, for simplicity, a quadratic energy density. More realistic (Ogden-type) expressions to explore the regime of large induced stresses are discussed in DeSimone and Teresi (2009) and Agostiniani and DeSimone (2011b). In fact, expression \eqref{energy} for the energy is the 1D, small strain version of the energy proposed by Warner, Terentjev and collaborators (Bladon {\it et al.}, 1993; Verwey {\it et al.}, 1996), and thoroughly discussed by DeSimone and coworkers in a series of papers (DeSimone, 1999; DeSimone and Dolzmann, 2000; DeSimone and Dolzmann, 2002; Conti {\it et al.}, 2002a; Conti {\it et al.}, 2002b; Agostiniani and DeSimone, 2011a).
The emergence of \eqref{energy} as the small strain limit of the Warner-Terentjev energy has been discussed on the basis of both formal Taylor expansion and Gamma-convergence arguments in DeSimone and Teresi (2009), Cesana and DeSimone (2011) and Agostiniani and DeSimone (2011b).
We remark that the model we are going to develop, based on energy \eqref{energy}, could be applied also to active strips made of other active materials (e.g., soft electroactive polymers, but also hard materials such as electrostrictive, ferroelectric, ferromagnetic, and ferroelastic solids). As we will see in the sequel, larger spontaneous strains lead to larger achievable displacements and locomotion is possible only  if the spontaneous strains are sufficiently large, in a sense made precise by inequalities \eqref{101} and \eqref{eq:C_emax} below. Since LCEs provide the key example of a soft active material exhibiting large spontaneous strains, they provide the most natural candidate material for which our model can deliver interesting results.

The tension $T(X,t)$ at any crawler section $X$ and time $t$ is given by
\bege\label{tension}
T(X,t)= K \left(  u^\prime(X,t)- \epso(X,t)   \right)= K u^\prime(X,t) + T^a(X,t) ,
\eend
and is the 1D analogue of the first Piola-Kirchhoff stress. The term $T^a(X,t)= -K \epso(X,t)$ can be regarded as the \emph{active} part of the internal tension, in analogy with the
\emph{active stress} used to model biological matter as an active gel (Marchetti {\it et al.}, 2013).
\begin{figure}[h]
\renewcommand{\figurename}{\small {\bf Fig.}}
\centering
\includegraphics{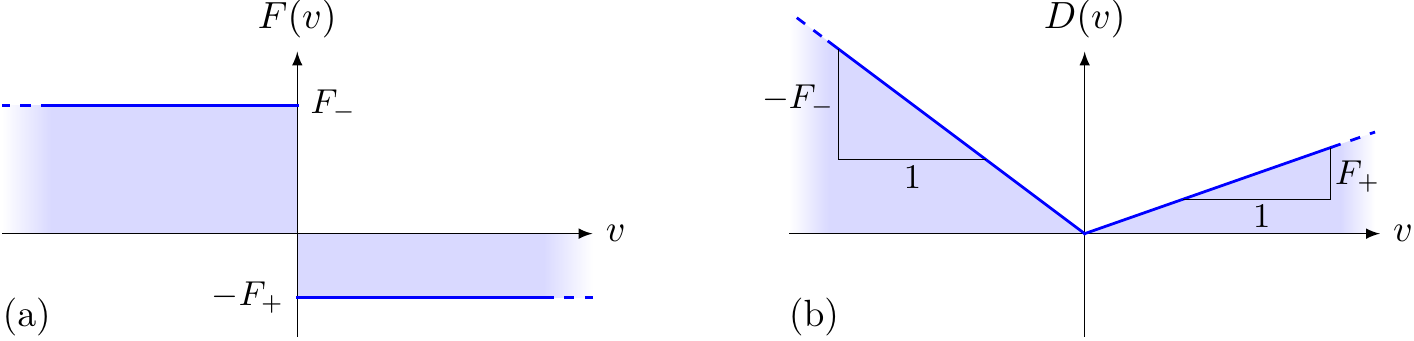}
\caption{\small Force-velocity law (a) and dissipation (b) in the case of frictional, directional forces acting only at the two extremities of the crawler.}
\label{fig:fric-conc}
\end{figure}

Frictional forces arising from directional interactions with a solid substrate act on the crawler. These are either concentrated at the two ends or distributed along the crawler body, see the sketch of Fig.~\ref{fig:crawlers}.

In the case of frictional forces acting only at the two ends of the crawler $X_i$ ($i=1,2$), these are given by
\bege
F_i(t)=F(\dot{u}_i(t)) , \text{ where } F(v) \in
\begin{cases}
\displaystyle \{F_-\}   &   \!\!\! \text{ if } v<0 , \\
\displaystyle [-F_+,F_-]   &   \!\!\! \text{ if } v=0 , \\
\displaystyle \{-F_+\}   &   \!\!\! \text{ if } v>0 ,
\end{cases}
\eend
and $F_->F_+>0$ are threshold forces to be overcome for sliding to occur to the left or to the right, respectively, see Fig.~\ref{fig:fric-conc}a. The assumption $F_->F_+$ simply means that we have chosen to orient the $x$-axis so that the positive direction is the one of easy sliding. We remark that the notation $F(v)\in\{F_-\}$ means $F(v)=F_-$, which occurs if $v<0$. Likewise, the notation $F(v)\in\{-F_+\}$ means $F(v)=-F_+$, which occurs if $v>0$. If instead $v=0$, then $F(v)$ can take any value in the interval $[-F_+,F_-]$. We also notice that the contribution of the end frictional forces to the rate of energy dissipation reads
\bege
-\sum_{i=1}^2 F_i(t) \dot{u}_i(t) = \sum_{i=1}^2 D(\dot{u}_i(t)) ,
\eend
where the dissipation $D(v) \defeq F_+\left( v \right)^+ - F_-\left( v \right)^-$ has been introduced with $\left( v \right)^\pm \defeq \frac{1}{2} \left(v \pm |v| \right)$,
such that in the notation of convex analysis we can write
\bege\label{one}
-F(v)\in \frac{\partial}{\partial v} D(v)=
\begin{cases}
\displaystyle \{-F_-\} & \!\!\! \text{ if } v<0 , \\
\displaystyle [-F_-,F_+] & \!\!\! \text{  if } v=0 , \\
\displaystyle \{F_+\} & \!\!\! \text{ if } v>0 , 
\end{cases}
\eend
where $\tfrac{\partial}{\partial v} D(v)$ is the sub-differential of $D$ at $v$, see Fig.~\ref{fig:fric-conc}b.

\begin{figure}[h]
\renewcommand{\figurename}{\small {\bf Fig.}}
\centering
\includegraphics{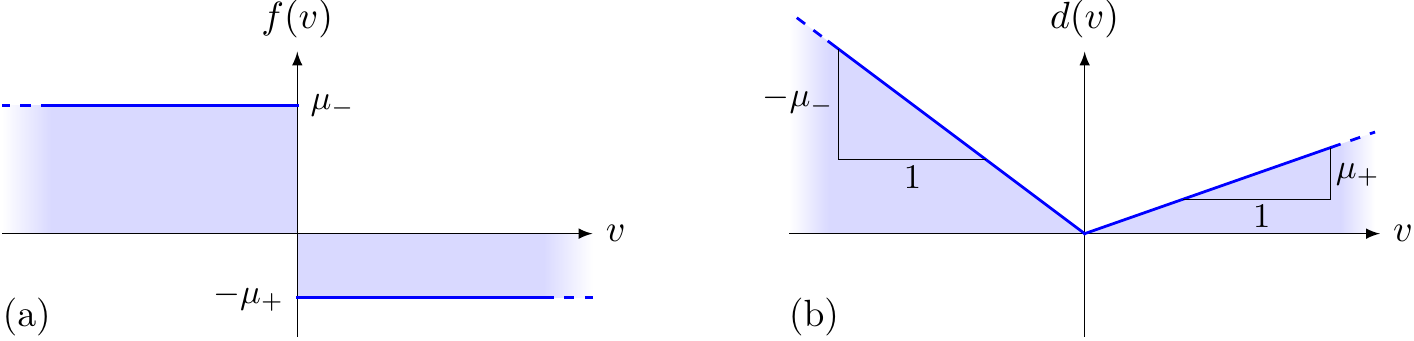}
\caption{\small Force-velocity law (a) and dissipation (b) in the case of frictional, directional forces distributed along the crawler body.}
\label{fig:fric-dist}
\end{figure}

In the case of distributed interactions, the frictional force per unit reference length is given by 
\bege
f(X,t)=f(\dot{u}(X,t))\,, \text{ where }f(v)\in
\begin{cases}
\displaystyle \{\mu_-\}  &   \!\!\! \text{ if } v<0 , \\
\displaystyle [-\mu_+,\mu_-]  &   \!\!\! \text{ if } v=0 , \\
\displaystyle \{-\mu_+\}  &   \!\!\! \text{ if } v>0 ,
\end{cases}
\eend
and $\mu_->\mu_+>0$ are threshold forces per unit reference length to be overcome for sliding to occur, see Fig.~\ref{fig:fric-dist}a. As before,  $f(v)\in\{\mu_-\}$ means $f(v)=\mu_-$, which occurs if $v<0$. Likewise, $f(v)\in\{-\mu_+\}$ means $f(v)=-\mu_+$, which occurs if $v>0$. If instead $v=0$, then $f(v)$ can take any value in the interval $[-\mu_+,\mu_-]$. The contribution of the distributed frictional forces to the rate of energy dissipation is now
\bege
-\int_{0}^L  f(X,t) \dot{u}(X,t)\dd X= \int_{0}^L  d(\dot{u}(X,t)) \dd X ,
\eend
where the dissipation per unit reference length $d(v) \defeq \mu_+\left( v \right)^+ - \mu_-\left( v \right)^-$ has been introduced, again with $\left( v \right)^\pm:= \frac{1}{2} \left( v \pm |v| \right)$,
such that in the notation of convex analysis we can write
\bege\label{two}
-f(v)\in \frac{\partial}{\partial v} d(v)=
\begin{cases}
\displaystyle \{-\mu_-\}  &   \!\!\! \text{ if } v<0 , \\
\displaystyle [-\mu_-,\mu_+]  &   \!\!\! \text{ if } v=0 , \\
\displaystyle \{\mu_+\}  &   \!\!\! \text{ if } v>0 ,
\end{cases}
\eend
where $\tfrac{\partial}{\partial v} d(v)$ is the sub-differential of $d$ at $v$, see Fig.~\ref{fig:fric-dist}b.

We consider a history of active distortions $\epso(X,T)$ varying in time sufficiently slowly, so that the crawler evolves quasi-statically through a sequence of equilibrium states. The governing equations are then obtained by neglecting inertia in the balance of linear momentum, and read
\bege\label{balance}
T^\prime (X,t) + f(\dot{u}(X,t)) = 0 ,
\eend
together with the boundary conditions for the tension at the crawler extremities
\bege\label{bc1}
\left\{
\begin{array}{l}
\!\! T(0,t)=-F_1(t)=-F(\dot{u}_1(t)) , \\[2mm]
\!\! T(L,t)=F_2(t)=F(\dot{u}_2(t)) .
\end{array}
\right .
\eend

Our quasi-static approximation may need to be reconsidered in some applications, where stick-slick phenomena may lead to oscillations, or  even in the interest of exploring dynamic effects that may lead to additional locomotion mechanisms. This occurs, for example, in the case liquid drops moving on a  vibrated substrate where the complex shape dynamics of the drop may lead to  reversal  of the direction of motion as the frequency and amplitude of vibration of the substrate are varied (Chaudhury {\it et al.}, 2015). It has been suggested in Cicconofri and DeSimone (2015) that a similar effect can also occur in bristle-legged-robots locomoting on a rigid substrate when actuated by rotary motors or by a vibrating internal mass.

By making use of \eqref{one} and \eqref{two}, the balance of linear momentum can be rewritten as
\bege\label{001}
T^\prime(X,t)\in \frac{\partial}{\partial v} d\left(\dot{u}(X,t)\right) ,
\eend
whereas the boundary conditions at the two extremities become
\bege\label{bc2}
\left\{
\begin{array}{l}
\displaystyle \!\! T(0,t)\in  \frac{\partial}{\partial v} D \left(\dot{u}_1(t)\right) , \\[2.5mm]
\displaystyle \!\! T(L,t)\in - \frac{\partial}{\partial v} D \left(\dot{u}_2(t)\right) .
\end{array}
\right .
\eend

Integrating \eqref{balance}, and using the boundary conditions \eqref{bc1}, we obtain the global force balance for the crawler, namely
\bege\label{004}
F(\dot{u}_1(t))+ F(\dot{u}_2(t))+ \int_0^L f (\dot{u} (X,t))\dd X =0\,.
\eend

\section{Formulation of the motility problem}\label{sec:MotilityPb}

\begin{figure}[h]
\renewcommand{\figurename}{\small {\bf Fig.}}
\centering
\includegraphics{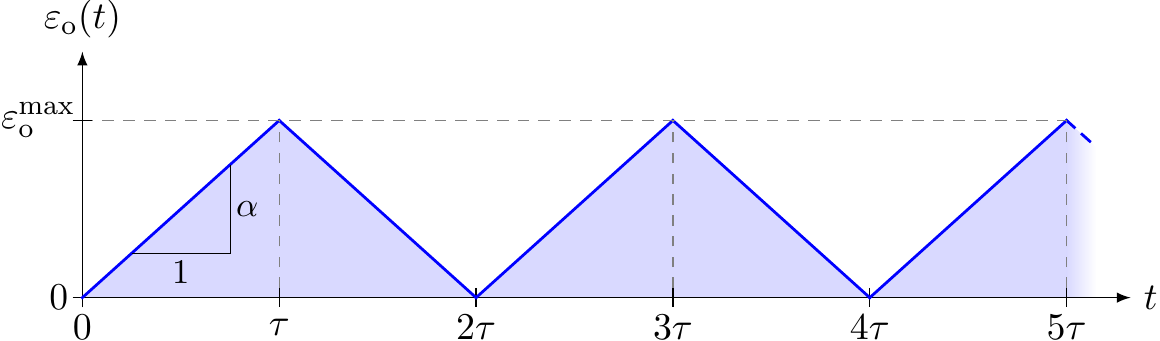}
\caption{\small Time history of $2\tau$-periodic, sawtooth active distortions applied to the crawler.}
\label{fig:dist-hist}
\end{figure}

We formulate our motility problem as follows. Given the initial state of the system through the assignment of the initial position and tension, e.g.,  $u(X,0)\equiv 0$, and $T(X,0)\equiv 0$, find the history of displacements $t \mapsto u(X,t)$ and tensions $t \mapsto T(X,t)$ corresponding to a given periodic time history of spatially constant active distortions, $t\mapsto \epso (X,t) \equiv \epso(t)$. In particular, find the asymptotic average speed of the crawler
\bege
\lim_{t \to+\infty}\frac{u(X^*, t)}{t} ,
\eend
where $X^*$ is an arbitrarily chosen point. We consider in particular the time history of active distortions given by the $2\tau$-periodic sawtooth graph of Fig.~\ref{fig:dist-hist}, defined on $[0,2\tau]$ as
\begin{equation}
\epso(t)=
\begin{cases}
\keps t		&	\text{for $t \in [0,\tau]$} , \\
\keps (2\tau-t)	&	\text{for $t \in [\tau,2\tau]$} ,
\end{cases}
\end{equation}
and then extended $2\tau$-periodically for $t \geq 0$. We denote the maximum distortion encountered as
\begin{equation}
\epsomax \defeq \keps \tau .
\end{equation}
Clearly, more general time histories could be of interest such as, for example, the travelling contraction waves considered in DeSimone and Tatone (2012), DeSimone {\it et al.} (2013) and Noselli {\it et al.} (2013), but we postpone the analysis of these more general cases to future work.
\subsection{Friction only at the ends}

Here, since $f\equiv0$, we have that the internal tension $T(X,t)$ is independent of the coordinate $X$. It follows from \eqref{tension} that also $u^\prime(X,t)$ is independent of $X$, and the expression \eqref{energy} for the energy reduces to
\begin{equation}\label{energy_reduced}
\Ec_r(u_1(t),u_2(t),t) = \frac{KL}{2} \left(\frac{u_2(t) - u_1(t)}{L} - \epso(t) \right)^{\!\!2} \!.
\end{equation}
Furthermore, the evolution equations \eqref{bc2} simplify to

\bege\label{bc3}
\left\{
\begin{array}{l}
\displaystyle \!\! T(0,t) = -\frac{\partial}{\partial u_1} \Ec_r(u_1(t),u_2(t),t)  \in  \frac{\partial}{\partial v} D \left(\dot{u}_1(t)\right) , \\[3.5mm]
\displaystyle \!\! -T(L,t) = -\frac{\partial}{\partial u_2} \Ec_r(u_1(t),u_2(t),t)  \in \frac{\partial}{\partial v} D \left(\dot{u}_2(t)\right) ,
\end{array}
\right .
\eend
and, at times when sliding occurs, these can be written as equalities
\bege\label{bc4}
\left\{
\begin{array}{ll}
\displaystyle \!\! -\frac{\partial}{\partial u_1} \Ec_r(u_1(t),u_2(t),t)  =  \frac{\partial}{\partial v} D\left(   \dot{u}_1(t)\right)	&	\text{ if }	\dot{u}_1 \neq 0 , \\[3.5mm]
\displaystyle \!\! -\frac{\partial}{\partial u_2} \Ec_r(u_1(t),u_2(t),t) =  \frac{\partial}{\partial v} D\left(   \dot{u}_2(t)\right)  	&	\text{ if }	\dot{u}_2 \neq 0 .
\end{array}
\right .
\eend

Denoting now by
\bege
F^i_{fric} \defeq \frac{\partial }{\partial  \dot{u}_i} \sum_{j=1}^{2}D(\dot{u}_j(t)) \quad \text{and} \quad F^i_{el} \defeq \frac{\partial }{\partial  u_i}\Ec_r(u_1(t),u_2(t),t) 
\eend
the {\it frictional} and {\it elastic} forces at the i-th end (i=1,2), we recover the interpretation of the evolution equations \eqref{bc4} above as force balances at the two ends, on each of which the total force consists of an elastic and of a frictional contribution, namely
\bege
F^i_{el} + F^i_{fric} = 0 \,.
\eend
We recall that we are working in the quasi-static regime, hence neglecting inertial forces. Alternatively, by multiplying each of the equations above by $\dot{u}_i$ we obtain
\bege
-\frac{\partial}{\partial u_i} \Ec_r \, \dot{u}_i  =  F^i_{fric} \, \dot{u}_i \,,
\eend
and we can interpret the evolution equations as the statement that the system evolves in such a way that the energy dissipation rate always matches the rate of release of elastic energy. 

\subsection{Only distributed friction}\label{subb}

Here there are no concentrated frictional forces at the two ends, recall the sketch of Fig.~\ref{fig:crawlers}, so that equations \eqref{bc2} simply reduce to
\bege\label{tension_bc}
\left\{
\begin{array}{l}
\displaystyle \!\! T(0,t)=0 \,, \\[2.5mm]
\displaystyle \!\! T(L,t)=0 \,,
\end{array}
\right .
\eend
and provide the boundary conditions for the tension field
\bege\label{aaa}
T(X,t)=K(\up(X,t)-\epso(t)) ,
\eend
which satisfies the evolution equation \eqref{001}, namely,
\bege
\Tp(X,t)\in
\begin{cases}
\{-\mu_-\}  &   \!\!\! \text{ if } \ud(X,t) < 0 , \\
[-\mu_- , \mu_+]  &   \!\!\! \text{ if } \ud(X,t)=0 , \\
\{\mu_+\}  &   \!\!\! \text{ if } \ud(X,t) > 0 . 
\end{cases}
\label{tension_ev}
\eend
By substituting \eqref{aaa} into \eqref{tension_ev}, we see that, in this case, the evolution equations take the form of a differential inclusion for the displacement field $u(X,t)$.

\section{Friction only at the ends}\label{sec:d1}

We solve in this section the evolution problem for the case in which frictional forces act only at the two ends. This can be considered as a warm up for the more difficult case in which distributed frictional forces act along the crawler body.

\subsection{Evolution equations}\label{sec:d11}

We recall that, in this case, the internal tension $T(X,t)$ is independent of $X$ and given by
\bege
T(t) = K \left( \frac{u_2(t)-u_1(t)}{L} - \epso(t) \right) .
\eend
The equations governing the evolution of the system are \eqref{bc3}, and they can be conveniently recast as
\bege
\dot u_1(t)=0 , \,\,\,\, \dot u_2(t)=0 , \quad \text{ if }\,\,\,
\begin{cases}
\displaystyle T(t) \in  \,\, ]\!-\!F_+,F_+[ \,, \\
\displaystyle T(t) = F_+ \, \text{ and } \, \epsodot (t) \ge 0 , \\
\displaystyle T(t) = - F_+ \, \text{ and } \, \epsodot (t) \le 0 ,
\end{cases}
\label{eq:DEV_0}
\eend
corresponding to the case of stationarity of the two crawler extremities, 
\bege
\dot u_1(t)=\keps L , \,\,\,\, \dot u_2(t)=0 , \quad \text{ if }\,\,\,\,\, T(t) = F_+ \, \text{ and } \, \epsodot (t) < 0 ,
\label{eq:DEV_1}
\eend
corresponding to the case of slip for the left hand side of the crawler and stationarity of the other one, and finally
\bege
\dot u_1(t)=0 , \,\,\,\, \dot u_2(t)=\keps L , \quad \text{ if }\,\,\,\,\, T(t) = -F_+ \, \text{ and } \, \epsodot (t) > 0 ,
\label{eq:DEV_2}
\eend
corresponding to the case of stationarity for the left hand side of the crawler and slip of the other one.

\subsection{Solution of the motility problem}\label{sec:d12}

We recall that the initial conditions are $u_1(0)=u_2(0)=0$ and $T(0)=0$, and we consider the case of sufficiently large distortion, namely we assume that
\begin{equation}\label{101}
\epsomax > \frac{2F_+}{K} .
\end{equation}

We will show that the motion of the crawler is characterized by a preliminary transient phase for $t\in[0,\tau]$, followed by a $2\tau$-periodic behaviour for $t>\tau$, with a constant
forward displacement of the crawler in each period.
An important role in our analysis will be played by the time constant
\begin{equation}
\td=\frac{2 F_+}{K\keps} ,
\end{equation}
and we notice that our assumption \eqref{101} is equivalent to $\td < \tau$.

\paragraph{Interval $0 \leq t < \td/2$\,.}
During this time interval $\epsodot(t)=\keps>0$ and $T(t)>-F_+$, so that we are in case \eqref{eq:DEV_0}. The two ends of the crawler are stationary ($\ud_1(t)=\ud_2(t)=0$), such that
\bege
u_1(t) = u_2(t) = 0 ,
\eend
and the tension in the crawler varies linearly in time as
\bege
T(t)=-K\keps t ,
\eend
reaching the critical value $T(\td/2)=-F_+$ at the end of the interval.

\paragraph{Interval $\td/2 \leq t < \tau$\,.}
In this time interval we still have $\epsodot(t)=\keps>0$, but now $T(t)=-F_+$, so we are in situation \eqref{eq:DEV_2}. The first end is stationary ($\ud_1(t)=0$) while the second one moves keeping the tension constant ($\ud_2(t)=\keps L$), leading to
\bege
u_1(t)=0 , \,\,\,\, u_2(t)=\keps L\left(t - \frac{\td}{2}\right) .
\eend
At the end of the time interval we have that $T(\tau)=-F_+$ and
\bege
u_1(\tau)=0 , \,\,\,\, u_2(\tau)=\keps L\left(\tau - \frac{\td}{2}\right) . 
\label{eq:Dpos_tau}
\eend

\paragraph{Interval $\tau \leq t < \tau+\td$\,.}
During this time interval $\epsodot(t)=-\keps<0$ and $T(t)<F_+$, so we are again in situation \eqref{eq:DEV_0}. The two ends are stationary ($\ud_1(t)=\ud_2(t)=0$) and therefore at time $t=\tau + \td$ the position of the crawler is given by \eqref{eq:Dpos_tau}. The tension instead increases linearly according to
\bege
T(t)=-F_+ +K\keps \left(t-\tau \right) ,
\eend
reaching at the end of this time interval the other critical value $T(\tau+\td)=F_+$.

\paragraph{Interval $\tau+\td\leq t < 2\tau$\,.}
In this time interval we still have $\epsodot(t)=-\keps<0$, but now $T(t)=F_+$, so we are in situation \eqref{eq:DEV_1}. The second end is stationary ($\ud_2(t)=0$) while the first one moves keeping the tension constant ($\ud_1(t)=\keps L$), leading to
\bege
u_1(t)=\keps L(t-\tau-\td) , \,\,\,\, u_2(t)=\keps L\left(\tau - \frac{\td}{2}\right) .
\eend
At the end of the time interval we have that $T(2\tau)=F_+$ and
\bege
u_1(2\tau)=\keps L(\tau-\td) , \,\,\,\, u_2(2\tau)=\keps L\left(\tau - \frac{\td}{2}\right) .
\label{eq:Dpos_2tau}
\eend

\paragraph{Interval $2\tau\leq t < 2\tau+\td$\,.}
During this time interval $\epsodot(t)=\keps>0$ and $T(t)>-F_+$, so that we are in case \eqref{eq:DEV_0}. The two ends are stationary ($\ud_1(t)=\ud_2(t)=0$) and so at $t=2\tau + \td$ the position of the crawler is still the one of \eqref{eq:Dpos_2tau}. The tension decreases linearly according to
\bege
T(t)=F_+ -K\keps (t-2\tau) ,
\eend
and reaches at the end of the time interval the critical value of $T(2\tau+\td)=-F_+$. In this time interval we observe a behaviour similar to that of the first interval $0<t<\td/2$, but in this case we have a greater initial tension ($F_+$ instead of $0$), so we need twice the time to reach the critical tension $-F_+$. 

\paragraph{Interval $2\tau+\td \leq t <3\tau$\,.}
In this time interval we still have $\epsodot(t)=\keps>0$, but now $T(t)=-F_+$, so we are in situation \eqref{eq:DEV_2}. The first end is stationary ($\ud_1(t)=0$) while the second one moves keeping the tension constant ($\ud_2(t)=\keps L$), leading to
\bege
u_1(t)=\keps L(\tau-\td) , \,\,\,\, u_2(t)=\keps L\left(\tau - \frac{\td}{2}\right)+\keps L(t-2\tau-\td) .
\eend
At the end of the time interval we have that $T(3\tau)=-F_+$ and
\bege
u_1(3\tau)=\keps L(\tau-\td) , \,\,\,\, u_2(3\tau)=\keps L\left(2\tau - \frac{3\td}{2}\right) .
\eend

\begin{figure}[h]
\renewcommand{\figurename}{\small {\bf Fig.}}
\centering
\includegraphics{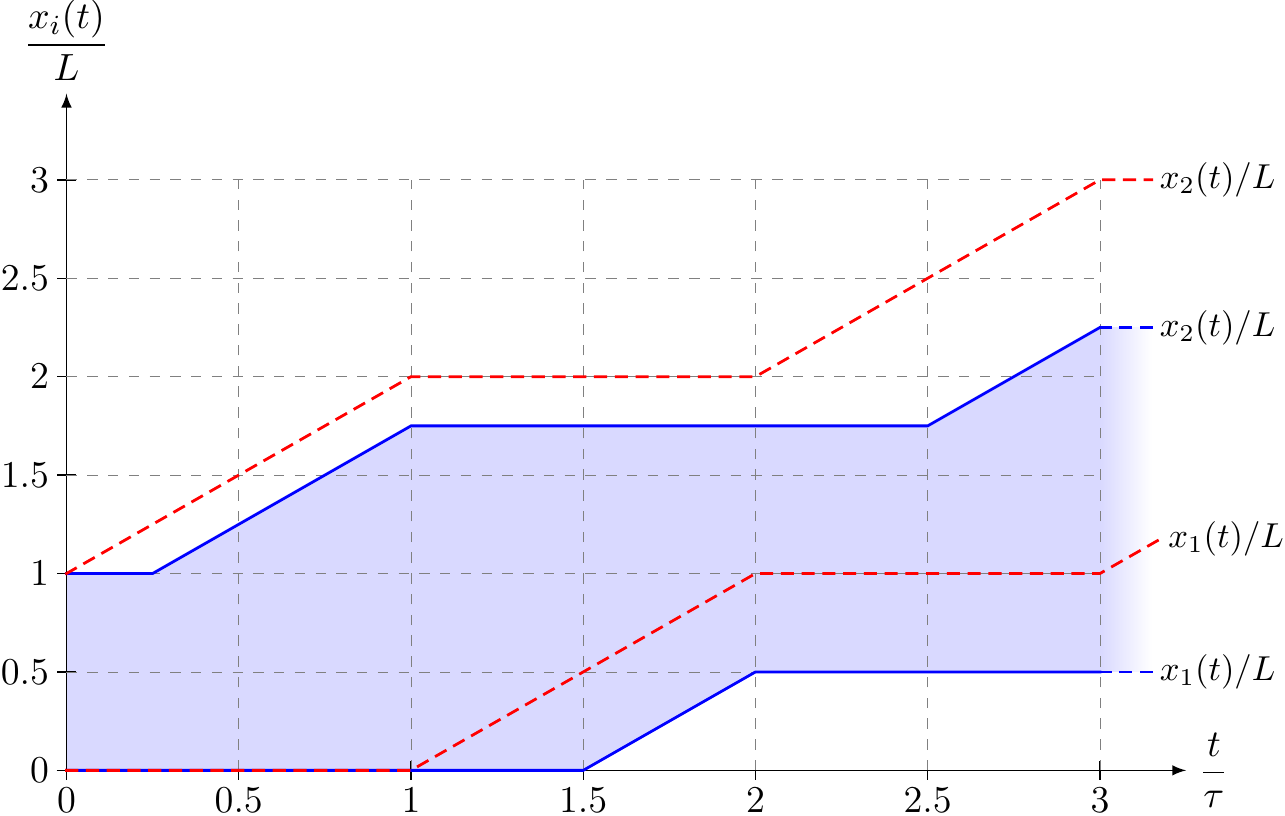}
\caption{\small Position of the crawler extremities $x_i(t)$ during a time interval of $3\tau$ and for a maximum distortion $\epsomax =1$. Two cases are shown to stress the effect on the displacements of the crawler stiffness, and these correspond to $\td/\tau = 1/2$ (or equivalently to $F_+/K=\epsomax/4$, blue solid curves), and $\td/\tau = 0$ (or equivalently to $F_+/K=0$, red dashed curves). Notice the piecewise linear time history of the displacements, which arises from the frictional, directional nature of the interactions with the substrate. At any time, the current length $l(t)$ of the crawler body (highlighted in the figure for the case of $\td/\tau = 1/2$) can be inferred from the vertical distance between the two curves.}
\label{fig:conc-adv}
\end{figure}

The position of the crawler extremities $x_i(t)$ is depicted in Fig.~\ref{fig:conc-adv} for a time interval of $3\tau$ and for the case of $\epsomax =1$. Specifically, two cases are shown to stress the effect on the displacements of the crawler stiffness, and these correspond to $\td/\tau = 1/2$ (blue solid curves), and $\td/\tau = 0$ (red dashed curves).
We observe that the state of the crawler at time $t=3\tau$ corresponds to that at time $t=\tau$ except for a translation of $\keps L(\tau-\td)$. Since the dynamics of the crawler is translation-invariant, the solution will repeat $2\tau$-periodically the behaviour found in $[\tau,3\tau]$. We can thus easily find the position of the crawler at any positive integer multiple of $\tau$, namely, for any integer $m>0$,
\bege
u_1(2m\tau)=m\keps L(\tau-\td) , \,\,\,\, u_2(2m\tau)=m\keps L(\tau-\td) +\keps L\frac{\td}{2} ,
\eend
and
\bege
u_1((2m+1)\tau)=m\keps L(\tau-\td) , \,\,\,\, u_2((2m+1)\tau)=(m+1)\keps L(\tau-\td) +\keps L\frac{\td}{2} .
\eend
Thus, the net displacement in one stretching cycle, corresponding to a time interval $\Delta t=2\tau$, reads 
\bege\label{engng}
\keps L(\tau-\td) = L \left(\epsomax -\frac{2F_+}{K}\right) .
\eend

The equation above shows that, at fixed $F_+$ and $K$, the achievable displacement increases when $\epsomax$ increases and no displacement is possible if the material exhibits spontaneous strains whose maximal magnitude does not satisfy inequality \eqref{101}.

Finally, we  notice that the crawler will be elongated in comparison to the initial length $L$, oscillating between a minimum length
\bege
l(2m\tau) = L \left(1+\frac{F_+}{K}\right) ,
\eend
and a maximum length
\bege
l((2m+1)\tau) = L \left(1+\epsomax -\frac{F_+}{K} \right) .
\eend

\section{Distributed friction}\label{sec:c1}

In the previous section, the case of a crawler has been addressed that exploits frictional, directional interactions at its ends only. We extend now our study to the case in which distributed frictional forces act along the crawler body.

\subsection{Evolution equations}\label{sec:c11}

We recall from Section \ref{subb} the evolution equations, namely,
\bege
\Tp(X,t)\in
\begin{cases}
\{-\mu_-\}  &   \!\!\! \text{ if } \ud(X,t) < 0 , \\
[-\mu_- , \mu_+]  &   \!\!\! \text{ if } \ud(X,t)=0 , \\
\{\mu_+\}  &   \!\!\! \text{ if } \ud(X,t) > 0 ,
\end{cases}
\label{eq:C_Tprime}
\eend
where the tension $T(X,t)$ is given by
\bege\label{Tdef}
T(X,t)=K(\up(X,t)-\epso(t)) .
\eend
It follows from equation \eqref{Tdef} that
\bege
\dot T(X_0,t)=-K\epsodot(t) \,\,\,\, \text{ if } \,\,\,\, \ud(X,t)=0 \,\,\,\,\, \forall  X \in N_0(X_0) ,
\label{eq:C_Tdot}
\eend
i.e., in a neighbourhood $N_0$ of $X_0$. Additionally, we have the boundary conditions at the crawler extremities, such that at any time
\bege\label{eq:C_bc}
\left\{
\begin{array}{l}
\displaystyle \!\! T(0,t)=0 \,, \\[2.5mm]
\displaystyle \!\! T(L,t)=0 \,.
\end{array}
\right .
\eend

\subsection{Solution of the motility problem}\label{sec:c12}

For the solution of the problem, it is expedient to introduce two special points, namely
\bege
X_L=\frac{\mu_+}{\mu_-  + \, \mu_+}L \,, \quad \, X_R=\frac{\mu_-}{\mu_- + \, \mu_+}L\,.
\eend
We first notice that
\bege
X_L+X_R=L \,,
\label{eq:C_L+R}
\eend
and set
\bege
x_L(t)=X_L+u_L(t) \,, \quad \, x_R(t)=X_R+u_R(t) \,,
\eend
where $u_L(t)=u(X_L, t)$ and $u_R(t)=u(X_R, t)$. We further notice that we can relate the positions at time $t$ of every couple of points $X_A$ and $X_B$ through 
\bege
u(X_B,t)=u(X_A,t)+\int_{X_A}^{X_B}\left(  \epso(t)+\frac1K T(X,t)  \right) \dd X ,
\label{eq:C_posAB}
\eend
which follows  from \eqref{Tdef}, by solving for $\up$ and then integrating with respect to $X$.

Similarly to the case of localized interactions, recall condition \eqref{101}, we assume that the active distortions are sufficiently large, and in fact require that
\bege
\epsomax > \frac{\mu_+}{K}L \,.
\label{eq:C_emax}
\eend

For our analysis, it is also useful to introduce two special time values, namely,
\bege
\tc = \frac{\mu_+ L}{K\keps} , \quad \, \tcc=\frac{\mu_-}{\mu_- + \, \mu_+}\frac{\mu_+ L}{K\keps} ,
\eend
and we notice that our assumption of large distortion is equivalent to $\tc < \tau$.

In what follows, we will seek solutions by using an ansatz on $\ud(X,t)$. Namely, we assume that the interval $[0,L]$ is partitioned into three, possibly empty, disjoint sub-intervals $I_L(t) \cup I_0(t) \cup I_R(t)=[0,L] $ (written in order from left to right) with either
\bege
\begin{cases}
\ud(X,t) <0 \quad \text{for } X\in I_L(t) ,\\
\ud(X,t) =0 \quad \text{for } X\in I_0(t) ,\\
\ud(X,t) >0 \quad \text{for } X\in I_R(t) ,
\end{cases}
\quad \text{ if } \,\, \epsodot(t) > 0 ,
\eend
i.e., for a positive incremental distortion, or
\bege
\begin{cases}
\ud(X,t) >0 \quad \text{for } X\in I_L(t) ,\\
\ud(X,t) =0 \quad \text{for } X\in I_0(t) ,\\
\ud(X,t) <0 \quad \text{for } X\in I_R(t) ,
\end{cases}
\quad \text{ if } \,\, \epsodot(t) < 0 , 
\eend
i.e., for a negative incremental distortion. We assume that $I_0(t)$ is a closed interval and consequently that $I_L(t)$ and $I_R(t)$ have an open end. The critical times $m\tau$, where $\epsodot$ is not defined, will be studied as extreme points of prescribed time sub-intervals and thus the partition of the ansatz will be assigned only  as (left or right) limit, in accordance with the instance considered.

Combining the ansatz with \eqref{eq:C_Tprime} and the boundary conditions \eqref{eq:C_bc} we deduce that, if $\epsodot (t)>0$, then
\bege
I_L(t)\subseteq [0,X_L[ \,, \quad I_R(t)\subseteq \, ]X_L, L] \,,
\eend
and the tension satisfies
\bege
\begin{cases}
T(X,t)=-\mu_- X & \text{if $X\in I_L(t)$} , \\
T(X,t)\geq -\mu_- X & \text{if $X\in I_0(t)\cap [0,X_L]$} , \\
T(X,t)\geq \mu_+(X-L) & \text{if $X\in I_0(t)\cap [X_L,L]$} , \\
T(X,t)=\mu_+(X-L) & \text{if $X\in I_R$} .
\end{cases}
\label{eq:C_Teplus}
\eend
In the two middle conditions of \eqref{eq:C_Teplus}, equality holds only on the boundary of $I_0$ in accordance with the continuity of $T$. On the other hand, for the interior points of $I_0$ the inequality is always strict, for else there would be a contradiction with \eqref{eq:C_Tdot}. In the extreme case $I_0(t)=\{X_L\}$,  the tension reaches everywhere its minimum admissible value 
\bege
\Tmin(X)=\begin{cases}
-\mu_- X & \text{if $0\le X \le X_L $} , \\
\mu_+(X-L) &  \text{if $X_L \le X  \le L$} ,
\end{cases}
\label{eq:C_Tmin}
\eend
and the whole crawler is extending, with each point moving away from the only stationary point $X_L$. We remark that, once this tension configuration is reached, we will have $T(X,t)=\Tmin(X)$ as long as $\epsodot(t)>0$, with the crawler elongating according to \eqref{Tdef}. In fact, any change in the tension would be in contradiction with \eqref{eq:C_Tprime}.

A similar reasoning is applicable in the case of a negative incremental distortion. In fact we argue that, if $\epsodot (t)<0$, then
\bege
I_L(t)\subseteq [0,X_R[ \, , \quad I_R(t)\subseteq \, ]X_R, L] \, ,
\eend
and the tension satisfies
\bege
\begin{cases}
T(X,t)=\mu_+ X & \text{if $X\in I_L(t)$} , \\
T(X,t)\leq \mu_+ X & \text{if $X\in I_0(t)\cap [0,X_R]$} , \\
T(X,t)\leq -\mu_-(X-L) & \text{if $X\in I_0(t)\cap [X_R,L]$} , \\
T(X,t)=-\mu_-(X-L) & \text{if $X\in I_R$} .
\end{cases}
\label{eq:C_Teminus}
\eend

As in the previous case, the inequalities are strict in the interior of $I_0(t)$, whereas the equality holds on the boundary of $I_0(t)$. In the limit case  $I_0(t)=\{X_R\}$,  the tension reaches everywhere its maximum admissible value
\bege
\Tmax(X)=\begin{cases}
\mu_+ X & \text{if $0\le X \le X_R $} , \\
-\mu_-(X-L) &  \text{if $X_R \le X  \le L$} ,
\end{cases}
\label{eq:C_Tmax}
\eend
and the whole crawler is contracting around the only stationary point $X_R$. The crawler will keep this tension configuration, i.e. $T(X,t)=\Tmax(X)$, as long as 
$\epsodot (t)<0$, contracting accordingly.

As in the case of friction only at the ends, we will show that the motion is characterized by a preliminary transient phase for $t\in[0,\tau]$, followed by a $2\tau$-periodic behaviour for $t>\tau$, with a constant forward displacement of the crawler in each period.
\begin{figure}[h]
\renewcommand{\figurename}{\small {\bf Fig.}}
\centering
\includegraphics{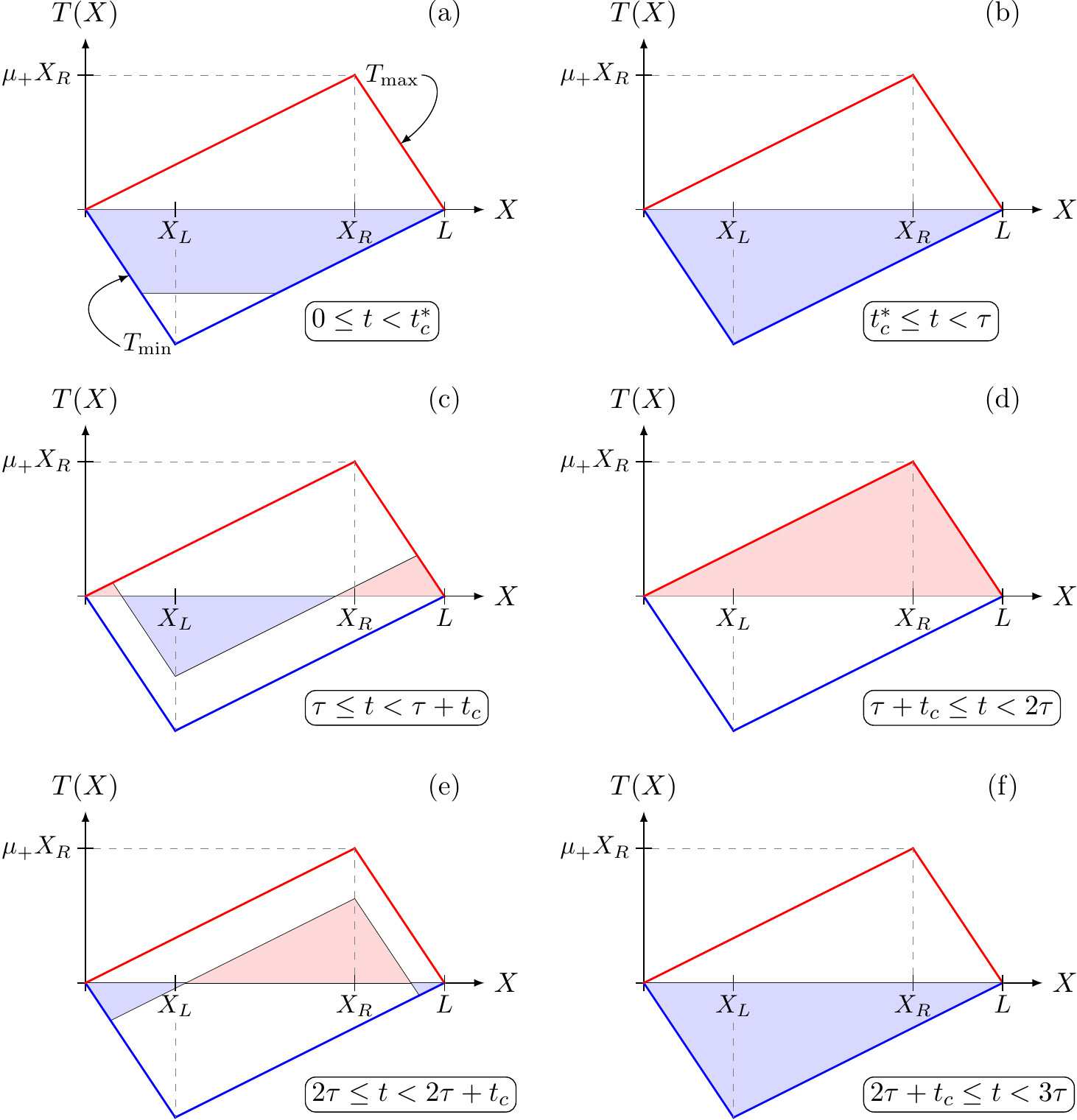}
\caption{\small Tension $T(X)$ along the crawler body during distinct time intervals for the case of distributed, directional friction. Notice that the tension stays always bounded and oscillates between the maximum and the minimum admissible values of $\Tmax$ and $\Tmin$, respectively.}
\label{fig:tensions}
\end{figure}

\paragraph{Interval $0\leq t< \tcc$\,.} 
We recall the initial conditions, namely $u(X,0) = 0$ and $T(X,0) = 0$. For \eqref{eq:C_Teplus}, at the beginning of the time interval the crawler is stationary, i.e.~$I_0(0)=[0,L]$, and so at every point $X$ the tension decreases according to \eqref{eq:C_Tdot}, until it reaches the critical value of $\Tmin(X)$, such that point $X$ starts to move, see Fig.~\ref{fig:tensions}a. Explicitly, we have that
\bege
T(X,t)=\begin{cases}
-\mu_- X		&	\text{if $0\le X < c_1 t$,  \quad i.e., if~$X\in I_L(t)$} , \\
-K \keps t		&	\text{if $c_1 t \leq X \leq L - c_2 t$, \quad i.e., if~$X\in I_0(t)$} , \\
\mu_+(X-L)	&	\text{if $L - c_2 t < X  \le L$,  \quad i.e., if~$X\in I_R(t)$} ,
\end{cases}
\eend
where the two velocities $c_1$ and $c_2$ have been introduced as
\bege
c_1 = \frac{K\keps}{\mu_-} , \quad c_2=\frac{K\keps}{\mu_+} .
\eend

At the end of the time interval, we have $T(X,\tcc)=\Tmin (X)$. We also notice that during the whole interval $X_L\in I_0(t)$, that means that $\ud_L\equiv 0$ and so
\bege
u_L(\tcc)=0 .
\eend
Hence, by using equations \eqref{eq:C_L+R} and \eqref{eq:C_posAB}, we obtain the expressions for the displacement of the crawler extremities at time $\tcc$
\bege
\left\{
\begin{array}{l}
\displaystyle \!\! u_1(\tcc) = -\keps \tcc X_L + \frac{\mu_-}{2K}X_L^2 , \\[3.5mm]
\displaystyle \!\! u_2(\tcc) = \keps \tcc  X_R - \frac{\mu_+}{2K}X_R^2 ,
\end{array}
\right .
\eend
and of point $X_R$, namely
\bege
u_R(\tcc) = \keps \tcc  (X_R-X_L) - \frac{\mu_+}{2K}(X_R^2-X_L^2) .
\eend

\paragraph{Interval $\tcc\leq t<\tau$\,.}
At time $t=\tcc$, the tension has reached its minimum value \eqref{eq:C_Tmin} everywhere along the crawler, and we still have $\epsodot (t)>0$ until the end of the interval, so, as we have anticipated, the tension remains constant, i.e. $T(X,t)=\Tmin(X)$, and the crawler elongates until $t=\tau$. Moreover, the point $X_L$ stands still and so
\bege
u_L(\tau)=u_L(\tcc)=0 .
\eend
Since the tension is known, see Fig.~\ref{fig:tensions}b, we can find the displacement of other points at time $t=\tau$ by comparison with $u_L(\tau)$ and using the condition \eqref{eq:C_posAB}. In this way, we immediately get the displacements of the extremities, namely
\bege
\left\{
\begin{array}{l}
\displaystyle \!\! u_1(\tau) = -\epsomax X_L +\frac{\mu_-}{2K}X_L^2 \,, \\[3.5mm]
\displaystyle \!\! u_2(\tau) = \epsomax X_R -\frac{\mu_+}{2K}X_R^2 \,,
\end{array}
\right .
\eend
and also of point $X_R$ 
\bege
u_R(\tau) = \epsomax (X_R-X_L) - \frac{\mu_+}{2K}(X_R^2-X_L^2) .
\eend

\paragraph{Interval $\tau\leq t<\tau+\tc$\,.}
During this time interval the crawler is subject to a negative incremental distortion, i.e. $\epsodot (t)<0$, so we are in the case \eqref{eq:C_Teminus}. At the beginning of the interval the crawler is stationary, and so the tension at each point $X$ increases according to \eqref{eq:C_Tdot}, until it reaches the maximum admissible value of $\Tmax(X)$, such that point $X$ begins to move, see Fig.~\ref{fig:tensions}c. Explicitly, we have now that
\bege
T(X,t)=\begin{cases}
\mu_+ X					&		\text{if $0\le X < c_3 (t-\tau)$} , \\
K \keps (t-\tau) -\mu_-X 		&		\text{if $c_3 (t-\tau) \leq X \leq X_L$} , \\
K \keps (t-\tau) +\mu_+(X-L) 	&		\text{if $X_L \leq X \leq L - c_3 (t-\tau)$} , \\
-\mu_-(X-L)				&		\text{if $L - c_3 (t-\tau) < X  \le L$} ,
\end{cases}
\eend
where the first and the last interval correspond respectively to $I_L(t)$ and $I_R(t)$, whereas the union of the other two is $I_0(t)$, and $c_3$ has been introduced as
\bege
c_3=\frac{K\keps}{\mu_- + \, \mu_+} .
\eend

At the end of the time interval we have $T(X,\tau+\tc)=\Tmax(X)$. Furthermore, we remark that during the whole interval we also have $[X_L,X_R]\subseteq I_0(T)$, since all the points of this subinterval reach the critical tension $\Tmax(X)$ simultaneously at $t=\tau+\tc$. It follows that $X_L$ and $X_R$ are stationary ($\ud_L\equiv \ud_R \equiv 0$) and thus
\bege
u_L(\tau+\tc) = u_L(\tau) , \quad u_R(\tau+\tc) = u_R(\tau) .
\eend
It turns out that the displacements of the two end points can be obtained from $u_L(\tau+\tc)$ and $u_R(\tau+\tc)$ by using \eqref{eq:C_posAB}, namely
\bege
\left\{
\begin{array}{l}
\displaystyle \!\! u_1(\tau+\tc) = -(\epsomax -\keps \tc )X_L -\frac{\mu_+}{2K}X_L^2 \,, \\[3.5mm]
\displaystyle \!\! u_2(\tau+\tc) = u_R(\tau+\tc)+(\epsomax -\keps \tc )X_L +\frac{\mu_-}{2K}X_L^2 \,.
\end{array}
\right .
\eend

\paragraph{Interval $\tau+\tc\leq t<2\tau$\,.}
At time $t=\tau+\tc$, the tension has reached its maximum value \eqref{eq:C_Tmax} everywhere along the crawler, and we still have $\epsodot (t)<0$ until the end of the interval, so the tension remains constant, i.e. $T(X,t)=\Tmax(X)$, and the crawler contracts until time $t=2\tau$. Furthermore, point $X_R$ stands still and so we immediately get
\bege
u_R(2\tau)=u_R(\tau+\tcc)=\epsomax (X_R-X_L) - \frac{\mu_+}{2K}(X_R^2-X_L^2) .
\eend
Since the tension is known, see Fig.~\ref{fig:tensions}d, we can find the displacement of other points at $t=2\tau$ by comparison with $u_R(2\tau)$ and using \eqref{eq:C_posAB}. In fact, the displacements of the extremities read
\bege
\left\{
\begin{array}{l}
\displaystyle \!\! u_1(2\tau) = u_R(2\tau) - \frac{\mu_+}{2K}X_R^2 \,, \\[3.5mm]
\displaystyle \!\! u_2(2\tau) = u_R(2\tau) + \frac{\mu_-}{2K} X_L^2 \,,
\end{array}
\right .
\eend
whereas for point $X_L$ we get
\bege
u_L(2\tau) = u_R(2\tau) - \frac{\mu_+}{2K}(X_R^2-X_L^2) = \epsomax (X_R-X_L) - \frac{\mu_+}{K}(X_R^2-X_L^2) .
\eend

\paragraph{Interval $2\tau \leq t<2\tau + \tc$\,.} 
During this time interval, the crawler is again subject to a incremental positive distortion, i.e. $\epsodot (t)>0$, and so we are in the case \eqref{eq:C_Teplus}. The crawler is stationary at the beginning of the interval, and consequently the tension at each point $X$ decreases according to \eqref{eq:C_Tdot}, until it reaches the minimum admissible value $\Tmin(X)$, such that point $X$ begins to move, see Fig.~\ref{fig:tensions}e. Explicitly, we have that
\bege
T(X,t)=\begin{cases}
-\mu_- X					& \text{if $0\le X < c_3 (t-2\tau)$} , \\
-K \keps (t-2\tau) +\mu_+X 	& \text{if $c_3 (t-2\tau) \leq X \leq X_R$} , \\
-K \keps (t-2\tau) - \mu_-(X-L) 	& \text{if $X_R \leq X \leq L - c_3 (t-2\tau)$} , \\
\mu_+(X-L) 				& \text{if $L - c_3 (t-2\tau) < X  \le L$} ,
\end{cases}
\eend
where the first and the last interval correspond to $I_L(t)$ and $I_R(t)$, respectively.

At the end of this time interval we have $T(X,2\tau+\tc)=\Tmin(X)$. We further underline that during the whole time interval we have $[X_L,X_R]\subseteq I_0(T)$. Specifically, points $X_L$ and $X_R$ are stationary during this interval ($\ud_L\equiv \ud_R \equiv 0$) and thus 
\bege
u_L(2\tau+\tc) = u_L(2\tau) , \quad u_R(2\tau+\tc) = u_R(2\tau) .
\eend
Again, the displacements of the end points can be conveniently computed by means of \eqref{eq:C_posAB}, namely
\bege
\left\{
\begin{array}{l}
\displaystyle \!\! u_1(2\tau+\tc) = u_L(2\tau+\tc)-\keps \tc X_L +\frac{\mu_-}{2K}X_L^2 \,, \\[3.5mm]
\displaystyle \!\! u_2(2\tau+\tc) = u_R(2\tau+\tc)+\keps \tc X_L -\frac{\mu_+}{2K}X_L^2 \,.
\end{array}
\right .
\eend

\paragraph{Interval $2\tau + \tc \leq t<3\tau$\,.} 
This time interval is qualitatively similar to the interval $\tcc\leq t<\tau$, but, since $\tcc <\tc$, it is shorter. We still have that $\epsodot (t)>0$, and the tension is constant in time and equals its minimum admissible value, i.e. $T(X,t)=\Tmin (X)$, see Fig.~\ref{fig:tensions}f. Therefore, the crawler elongates during this interval, with $X_L$ as a single stationary point, such that
\bege
u_L(3\tau) = u_L(2\tau+\tc) = \epsomax (X_R-X_L) - \frac{\mu_+}{K}(X_R^2-X_L^2) .
\eend
As for the previous time intervals, the displacements of the extremities can be easily computed by making use of \eqref{eq:C_posAB}, namely
\bege
\left\{
\begin{array}{l}
\displaystyle \!\! u_1(3\tau) = u_L(3\tau)-\epsomax X_L +\frac{\mu_-}{2K}X_L^2 \,, \\[3.5mm]
\displaystyle \!\! u_2(3\tau) = u_L(3\tau)+\epsomax X_R -\frac{\mu_+}{2K}X_R^2 \,,
\end{array}
\right .
\eend
whereas the displacement of point $X_R$ at time $t=3\tau$ reads
\bege
u_R(3\tau) = u_L(3\tau)+\epsomax (X_R-X_L) - \frac{\mu_+}{2K}(X_R^2-X_L^2) .
\eend
\begin{figure}[h]
\renewcommand{\figurename}{\small {\bf Fig.}}
\centering
\includegraphics{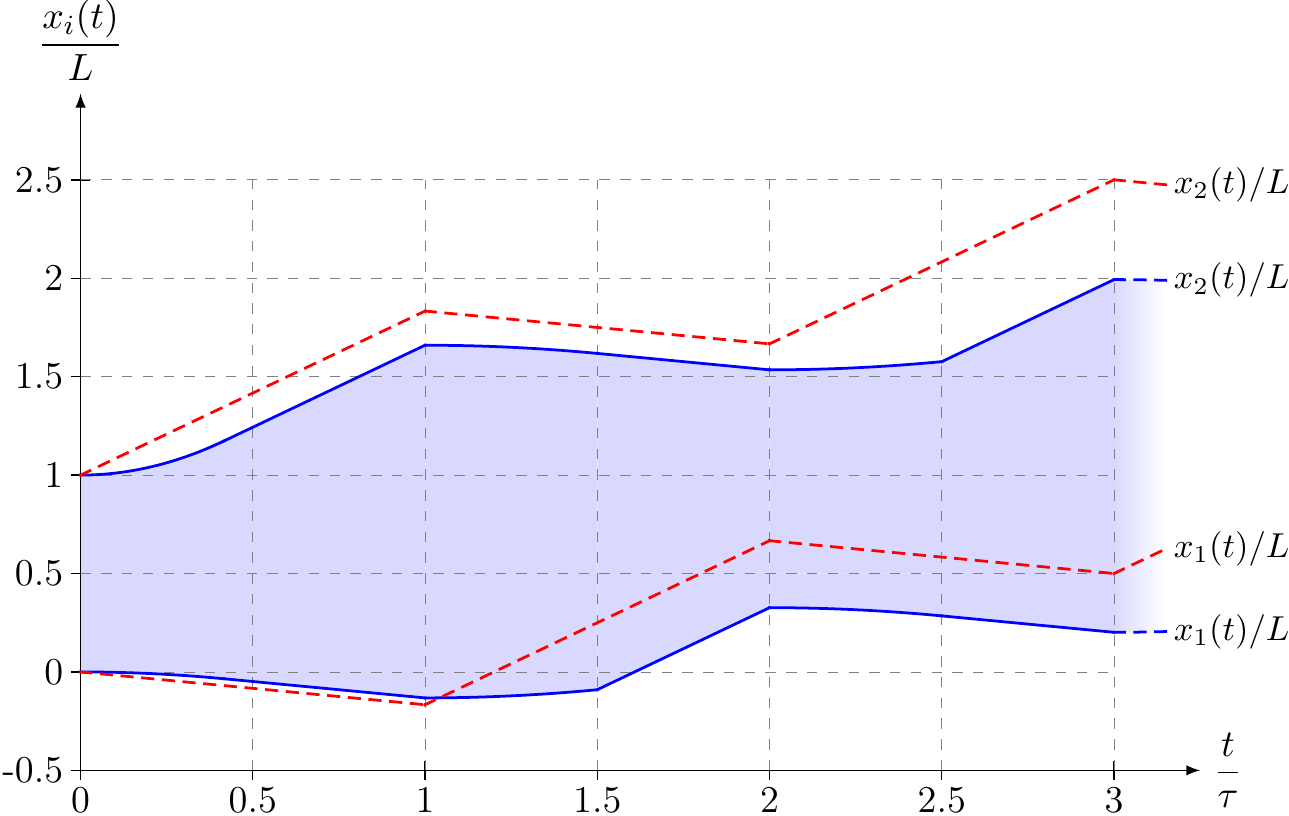}
\caption{\small Position of the crawler extremities $x_i(t)$ during a time interval of $3\tau$ for a maximum distortion $\epsomax =1$ and for a ratio of $\mu_+/\mu_- = 1/5$. Two cases are shown to stress the effect on the displacements of the crawler stiffness, and these correspond to $\mu_- L/K = 5/2$ (blue solid curves), and $\mu_- L/K = 0$ (red dashed curves). Notice that for the case of distributed friction, a significant back-sliding of $x_1$ takes place irrespective of the crawler stiffness $K$. At any time, the current length $l(t)$ of the crawler body (highlighted in the figure for the case of $\mu_- L/K = 5/2$) can be inferred from the vertical distance between the two curves.}
\label{fig:dist-adv}
\end{figure}

The position of the crawler extremities $x_i(t)$ is depicted in Fig.~\ref{fig:dist-adv} during a time interval of $3\tau$ for the case of $\epsomax =1$ and for a ratio of $\mu_+/\mu_- = 1/5$. Specifically, two cases are shown to stress the effect on the displacements of the crawler stiffness, and these correspond to $\mu_- L/K = 5/2$ (blue solid curves), and $\mu_- L/K = 0$ (red dashed curves).
We notice that the state of the crawler at time $t=3\tau$ corresponds to that at time $t=\tau $ except for a translation of $u_L(3\tau)-u_L(\tau)=u_L(3\tau)$. Since the dynamic of the crawler is translation-invariant, the behaviour found in the interval $[\tau,3\tau]$ will repeat $2\tau$-periodically. Furthermore, the asymptotic displacement produced in a $2\tau$-cycle can be reformulated as
\bege
u(X,3\tau)-u(X,\tau)=\left(\epsomax-\frac{\mu_+L}{K}\right)\frac{\mu_- - \mu_+}{\mu_- + \mu_+}L .
\label{eq:C_displacement}
\eend
To understand the meaning of the result above, we first recall that we have required the maximum distortion to satisfy the condition $\epsomax>\mu_+L/K$. In fact, it can be easily shown that, otherwise,   no net displacement can be extracted on average from periodic shape changes. 

The displacement \eqref{eq:C_displacement} produced in a cycle is actually linear with respect to the body length (and therefore scale invariant) if instead of the distortion $\epso$ we consider the distortion excess over the critical threshold of $\mu_+L/K$. The quadratic part of \eqref{eq:C_displacement} is only due to the fact that, keeping constant the other parameters, the distortion needed to produce some net motion linearly increases with the crawler length.
Finally, we observe that the length of the crawler oscillates between a minimum value that is reached at times that are even multiples of $\tau$, namely 
\bege
l(2m\tau)=L+\frac{\mu_-\mu_+ L^2}{2K(\mu_- +\mu_+)} ,
\eend
and a maximum value that, instead, is reached at times that are odd multiples of $\tau$, 
\bege
l((2m+1)\tau)=L+\left(\epsomax -\frac{\mu_-\mu_+ L}{2K(\mu_- +\mu_+)} \right) L .
\eend

\section{Discussion}
In this paper, a model crawler has been analysed that exploits frictional, directional interactions either concentrated at its extremities or distributed along the body.
Solutions to the motility problem have been obtained for both  cases. In particular, we provide explicit formulae for the net available displacements and for the tension acting at the crawler sections as a function of time, position and magnitude of the active distortions. These findings extend previous results by the authors. In fact, similarly to the approach followed in Gidoni {\it et al.} (2014), we consider the system as subject to a spatially uniform time-history of distortion, but now we do not assume the shape of the crawler to be known a-priori. Instead, the configuration of the crawler is an \emph{emergent} property which arises from the coupled nonlinear system consisting of the crawler force-generating mechanism, its passive elasticity and the external frictional forces. 
This approach has allowed us, in particular, to determine  the axial forces acting along the body of the crawler:  this is a quantity of great mechanical relevance in assessing the propensity of the system towards buckling, when compressions are generated during the locomotion process.
We plan to address this issue in future studies and to extend our approach to more complex time-histories of active distortions, such as those caused by the propagation of contraction waves.

While our analysis has focused mostly on some of the theoretical challenges that our model crawler raises, it already provides clues that may guide practical design.
For example, \eqref{engng}  implies that no net displacement can be achieved unless the available spontaneous strains are large enough that inequality~\eqref{101} is satisfied. Using order of magnitude estimates  for the geometric and material parameters involved, namely, $F_+=0.45\,$N (as in Noselli and DeSimone, 2014) and $K=10^2\,$N (arising from $K=EA$, with Young's modulus $E=1\,$MPa and area $A=10^{-4}\,$m$^2$ for a square cross-section of $1\,$cm\,$\times\,1\,$cm) we obtain that the minimal magnitude of the active strains to produce non-zero displacements is around $1\%$. In these circumstances, using Euler's formula for the buckling of a simply supported rod, one would estimate that strips can safely locomote without buckling provided that their length $L$ is below 10\,cm. Smaller scale cross sections (in particular, smaller thicknesses) will presumably require contact interactions with smaller $F_+$.

It is important to acknowledge that, while in our analysis we have taken for granted the existence of frictional interactions where the tangential force depends only on the sign of the slip velocity, with different values in the plus and minus direction, engineering such interactions in a practical device is far from trivial. The prototype crawlers described in 
Menciassi {\it et al.} (2006),  Noselli and DeSimone (2014) implement  directional frictional interactions by inserting asymmetric `microscopic feet' between the crawler body and the supporting substrate. But the question of which are the details of the frictional interaction laws emerging at the macroscopic level  from a given microscopic contact mechanism is very interesting, and largely open.
Among the issues worth addressing are the dependence of the emerging frictional forces on the roughness of the supporting surface, on  the normal forces acting at the points of microscopic contacts (due to gravity, or possibly to van der Waals interactions for devices of smaller size), static versus dynamic friction,  stick-slip phenomena, and dynamic effects.

We hope that our results may contribute to the development of effective limbless locomotion strategies in microscopic soft devices. In this area and, more generally, in the development of medical microscopic machines, LCEs may provide an interesting avenue towards miniaturization of artificial devices by exploiting the  principle \emph{the material is the machine} (Bhattacharya and James, 2005).  As suggested by the discussion above, the main challenges to be overcome are twofold. First and foremost,  making actual devices capable of implementing smart conceptual locomotion schemes. On the other hand, progress will also be needed in the development of new modelling and numerical tools  in order to tackle the nonlinear problems of motion planning and control for devices based on materials capable of large elastic deformations, since a general theory to handle them is still missing.

\section*{Acknowledgement}
This work has been supported by the ERC Advanced Grant 340685-MicroMotility. P.G. acknowledges also partial support from GNAMPA.

\appendix
\renewcommand\thesection{Appendix \Alph{section}}
\renewcommand\thesubsection{\Alph{section}.\arabic{subsection}}
\numberwithin{equation}{section}
\renewcommand{\theequation}{\Alph{section}.\arabic{equation}}

\section{Evolution equations through an incremental, variational principle}\label{sec:app}

The evolution equations solved in Sections~\ref{sec:d1} and \ref{sec:c1} can be obtained from an incremental variational principle, as we show below. This fact is not just a mathematical curiosity, because it could lead to the construction of solution algorithms similar to those that have proved successful in plasticity (Ortiz and Stainier, 1999; Miehe {\it et al.}, 2002), or in the study of the evolution of capillary drops subject to contact angle hysteresis phenomena (DeSimone {\it et al.}, 2007; Fedeli {\it et al.}, 2011; Alberti and DeSimone, 2011).

We consider the history of prescribed states of spontaneous distortions $t\mapsto \epso(t)$ given in Section \ref{sec:MotilityPb}, and assume that the displacement field $X\mapsto u(X,t)$ is known at time $t$. We look for the displacement and tension at time $t+\dd t$ by seeking solutions of the following incremental minimization problem. Find $X\mapsto u(X,t+dt)$ as
\bege\label{IncMinGen}
u(X, t+\dd t)= \argmin_{v}\left\{ \Ec(v,t+\dd t) +\diss(v, u(\cdot,t)) + \Diss(v, u(\cdot,t))  \right\} ,
\eend
where $\Ec$ is the elastic energy defined in \eqref{energy} and
\bege
\diss(v, u(\cdot,t)) \defeq \int_0^L \left\{   \mu_+\left( v(X)- u(X,t)  \right)^+ - \mu_-\left( v(X)- u(X,t)  \right)^- \right\} \dd X ,
\eend
whereas
\bege
\Diss(v, u(\cdot,t)) \defeq F_+\left( v_1-u_1(t) \right)^+ - F_-\left( v_1-u_1(t) \right)^- + F_+\left( v_2-u_2(t) \right)^+ - F_-\left( v_2-u_2(t) \right)^- \! .
\eend
Here we have set $v_1 \defeq v(X_1=0)$ and $v_2 \defeq (X_2=L)$. Once $u(X, t+\dd t)$ is known, we can find $T(X, t+\dd t)$ using \eqref{tension}, namely
\bege
T(X,t)= K \left(  u^\prime(X,t)- \epso(X,t)   \right) .
\eend

We consider the case of distributed friction first. We check for solutions of the form  $u(X, t+\dd t) = u(X,t)+ \dot u (X,t) \dd t$ and assume that $X \mapsto \ud (X,t)$ is continuous. 

Let us first prove \eqref{eq:C_Tprime} for every point $x_0\in (0,L)$ and for every time $t$ such that $\dot u (X,t) \ge 0$ in a neighbourhood $N_{x_0}^+$ of $x_0$. Using minimality of $u(X, t+\dd t)$ against $v_\eta(X) \defeq u(X,t)+ \dot u (X,t) \dd t + \eta \phi(X)$, with  $\eta \ge 0$ an arbitrary non-negative scalar, and $\phi(X) \ge 0$ an arbitrary non-negative $\mathcal{C}^\infty$ function with compact support in $N_{x_0}^+$, we obtain
\begin{flalign}
& \Ec(u(\cdot, t+\dd t), t+\dd t) + \diss(u(\cdot, t)+\dot u (X,t) \dd t, u(\cdot,t)) \le \notag \\[1.5mm]
&\quad\quad\quad\quad \Ec(u(\cdot, t+\dd t) + \eta\phi(\cdot), t+\dd t) + \diss(u(\cdot, t)+\dot u (X,t) \dd t+\eta\phi(\cdot) , u(\cdot,t)) \, ,
\end{flalign}
and, in turn,
\bege
\displaystyle I_\phi(\eta) \defeq \Ec(u(\cdot, t+\dd t)+\eta \phi(\cdot), t+\dd t) - \Ec(u(\cdot, t+\dd t) t+\dd t) + \eta \int_{N_{x_0}^+} \mu_+\phi(x)\dd X \ge 0
\eend
for every $\eta\ge 0$ and $\phi\ge 0$. Moreover, $I_\phi(0)=0$ for every $\phi$. It follows that
\bege
\label{derivative}
\frac{\dd}{\dd\eta}I_\phi (\eta)|_{\eta=0^+}= \int_{N_{x_0}^+} [ -K(u^\prime -\epso)^\prime + \mu_+ ] \phi \dd X \ge 0 
\eend
for every $\phi$, and since we can take an arbitrarily small neighbourhood $N_{x_0}^+$, we obtain that, for every $x_0\in (0,L)$ with $\dot u(X,t)\ge 0$ in a neighbourhood $N_{x_0}^+$,
\bege
\label{01}
-T^\prime (x_0, t+\dd t) + \mu_+  \ge 0 \,.
\eend
If, in particular, $\dot u (x_0,t)>0$, then there exists a neighbourhood $N_{x_0}^+$ where $\dot u (X,t)>0$ and we can take $\eta$ of unrestricted sign in the argument above. This leads to strict equality to zero in \eqref{derivative} and hence
\bege
\label{02}
-T^\prime (x_0, t+\dd t) + \mu_+  = 0 \,, \text{ at any } x_0\in (0,L) \text{ with  } \dot u(x_0,t) > 0 \,.
\eend

Similar arguments at a point $x_0\in (0,L)$ such that either $\dot u(x_0,t)\le0$, or $\dot u(x_0,t)< 0$, show that, for every $x_0\in (0,L)$ with $\dot u(X,t)\le 0$ in a neighbourhood $N_{x_0}^-$,
\bege
\label{03}
-T^\prime (x_0, t+\dd t) - \mu_-  \le 0 \,,
\eend
and that
\bege
\label{04}
-T^\prime (x_0, t+\dd t) - \mu_-  = 0 \,, \text{ at any } x_0\in (0,L) \text{ with  } \dot u(x_0,t) < 0 \,.
\eend

We exclude from our analysis the points where $\ud$ changes sign. Since $\ud$ is continuous, those point are at most countably many and thus negligible. Hence, putting \eqref{01}-\eqref{04} together, and using \eqref{two}, we obtain \eqref{001}, namely,
\bege
\label{a}
T^\prime(X,t)\in \frac{\partial}{\partial v} d\left(\dot{u}(X,t)\right) .
\eend

Now we derive the boundary conditions \eqref{eq:C_bc}. Given our solution $u(X, t+\dd t) = u(X,t)+ \dot u (X,t) \dd t$, we define $N^+ = \{X\in [0,L] \colon \ud(X,t)>0\}$ and $N^- = \{X\in [0,L] \colon \ud(X,t)<0\}$. Let $\phi(X)\geq 0$ be a non-negative $\mathcal{C}^\infty$ function on $[0,L]$ such that $\phi(L)>0$ and $\phi(0)=0$.
For every $\eta$ we set 
\bege
\A(\eta)=\begin{cases}
\{X\in N^+ \colon \ud(X,t)\dd t+\eta \phi(X)<0\} &\text{if $\eta<0$} ,\\
\emptyset &\text{if $\eta=0$} ,\\
\{X\in N^- \colon \ud(X,t)\dd t+\eta \phi(X)>0\} &\text{if $\eta>0$} .
\end{cases}
\eend
We have that $\abs{\A(\eta)}\to 0$ for $\eta\to 0$. We repeat the minimality argument used previously and obtain
\begin{flalign} 
I_\phi(\eta) \defeq \Ec(u(\cdot, t+\dd t)+\eta \phi(\cdot), t+\dd t) & - \Ec(u(\cdot, t+\dd t) t+\dd t) \, + \\[1.5mm]
& + \eta \int_{N_{L}^+} \mu_+\phi(x)\dd X + \eta \int_{N_{L}^-} \mu_-\phi(x)\dd X +\Rc(\eta) \ge 0 ,
\end{flalign}
where 
\bege
\Rc(\eta)=
\int_{\A (\eta)} (\mu_-+\mu_+)\abs{\ud(X,t)\dd t+\eta \phi(X)}\dd X .
\eend
We observe that
\bege
\abs{\Rc(\eta)}\le \abs{\eta}\abs{\A(\eta)}\abs{(\mu_-+\mu_+)\max_{X\in[0,L]}\phi(X)} ,
\eend
from which it follows that
\bege
\frac{\dd}{\dd\eta}\Rc (\eta)|_{\eta=0} = 0 .
\eend
From this condition and the minimality of $I_\phi(0)$ we obtain
\begin{flalign}
0 = \frac{\dd}{\dd\eta}I_\phi (\eta)|_{\eta=0} & = 
\int_{N_{L}^+} [-T(X,t+\dd t)+\mu_+]\phi(x)\dd X \, + \\[1.5mm]
& + \int_{N_{L}^-} [-T(X,t+\dd t)-\mu_-]\phi(x)\dd X +T(L)\phi(L)-T(0)\phi(0) .
\end{flalign}

The two integrals are both equal to zero because the integrands vanish in view of \eqref{02} and \eqref{04}. Since $\phi(L)=0$ and $\phi(0)>0$, we get $T(0)=0$. To obtain the second boundary condition $T(L)=0$ it suffices to consider instead test functions $\phi(X)\ge 0$ such that $\phi(L)>0$ and $\phi(0)=0$.

We now consider the case of friction concentrated at the two ends. Since $d \equiv 0$ implies that $T(X,t)$ is now independent of $X$ and, since $\epso$ is spatially uniform, the function  $X \mapsto u(X,t)$ is affine and the incremental minimization problem 
\eqref{IncMinGen} can be reduced to
\bege
\label{IncMinDisc}
\ub(t+\dd t)= \argmin_{\vb}\left\{ \Ec_{r}(\vb,t+\dd t) +\Diss(\vb, \ub(t) ) \right\} ,
\eend
where $\ub (t) \defeq (u_1(t), u_2(t))$ and $\vb (t) \defeq (v_1(t), v_2(t))$, whereas
\bege
\label{energy_red}
\Ec_r(\vb,t) \defeq \frac12 KL \left(\frac{v_2 -v_1}{L} -\epso (t)\right)^2 \!,
\eend
and
\bege
\Diss(\vb, \ub(t)) = F_+\left( v_1-u_1(t) \right)^+ - F_-\left( v_1-u_1(t) \right)^-  + F_+\left( v_2-u_2(t) \right)^+ - F_-\left( v_2-u_2(t) \right)^- \! . 
\eend

Following similar arguments to those used above for the case of distributed friction, we obtain \eqref{bc3}, namely,
\bege\label{b}
\left\{
\begin{array}{l}
\displaystyle \!\! T(0,t) =  - \frac{\partial}{\partial u_1} \Ec_r (\ub(t),t)  \in  \frac{\partial}{\partial u_1} D\left(   \dot{u}_1(t)\right) , \\[3.5mm]
\displaystyle \!\! -T(L,t)=  - \frac{\partial}{\partial u_2} \Ec_r (\ub(t),t) \in  \frac{\partial}{\partial \dot u_2} D\left(   \dot{u}_2(t)\right) .
\end{array}
\right .
\eend

We remark in conclusion that the general case in which both distributed and concentrated frictional forces are present can be obtained by combining equations \eqref{a} and \eqref{b}.

\end{document}